\documentstyle[12pt,twoside,fleqn,espcrc1,psfig]{article}

\newcommand{\doe}
{This work was supported by the
Director, Office of Energy Research,
Office of High Energy
and Nuclear Physics,
Division of Nuclear Physics,
of the U.S. Department of Energy under Contract
DE-AC03-76SF00098.}

\newcommand{\msun}{M_{\odot}}
\newcommand{\eos}{equation of state~}

\newcommand{\eosp}{equation of state}

\newcommand{\Eoss}{Equations of state}

\newcommand{\GR}{General Relativity~}

\newcommand{\beqn}{\begin{eqnarray}}
\newcommand{\eeqn}{\end{eqnarray}}

\newcommand{\tit}
{Pulsar Signal of Deconfinement}
\newcommand{\autha} {Norman K. Glendenning}
\newcommand{\dateofdoc}{16 December 1997}
\newcommand{\adra}
{Nuclear Science Division \&
Institute for Nuclear and Particle Astrophysics,
  Lawrence Berkeley Laboratory,
   MS: 70A-3307, Berkeley, California 94720}

\newcommand{\AmS}{{\protect\the\textfont2
  A\kern-.1667em\lower.5ex\hbox{M}\kern-.125emS}}

\title{Pulsar Signal of Deconfinement}

\author{Norman K. Glendenning\address{Nuclear Science Division \&
Institute for Nuclear and Particle Astrophysics,\\
  Lawrence Berkeley National Laboratory,
   MS: 70A-3307,\\ Berkeley, California 94720}%
        \thanks{\doe}}

\begin{document}

\begin{titlepage}
\parbox{4.5in}{\begin{flushleft}Quark Matter 97,
Tsukuba, Japan, (Dec. 1-5, 1997) %
\end{flushleft}}%
\parbox{1.5in}{\begin{flushright} LBNL-41164 \end{flushright}}
\begin{center}
\begin{Large}
\renewcommand{\thefootnote}{\fnsymbol{footnote}}
\setcounter{footnote}{1}
\tit {\footnote{\doe}}\\[5ex]
\end{Large}

\begin{large}
\autha\\[3ex]
\end{large}
\adra\\[3ex]
\dateofdoc \\[3ex]
\end{center}

\begin{figure}[htb]
\vspace{-.8in}
\begin{center}
\leavevmode
\hspace{-.7in}
\psfig{figure=ps.qm9,width=4in,height=2.66in}
\end{center}
\end{figure}


\begin{quote}
\begin{center}
Plenary Talk at  the International  Conference on Ultra-Relativistic\\
Nucleus-Nucleus Collisions\\
 {\bf Quark Matter 1997},\\
 Tsukuba, Japan\\
 To be published in Nuclear Physics A\\
 (Organizers: K. Yagi and S. Nagamiya)
 \end{center}
 \end{quote}
                                                            
\end{titlepage}


\maketitle

\begin{abstract}
A solitary millisecond pulsar, if near the mass limit, and
undergoing a phase transition, either first or second order,
provided the transition is to a substantially more compressible phase,
will emit a blatantly obvious signal---spontaneous spin-up.  Normally
a pulsar spins down by angular momentum loss to radiation.
The signal is trivial to detect and  is estimated to be
``on'' for 1/50 of the spin-down era of millisecond pulsars.
 Presently about 25 solitary millisecond pulsars are known.
The phenomenon is analogous to ``backbending'' observed in high
spin nuclei in the 1970's.
\end{abstract}

\section{THE LIVES OF NEUTRON STARS}
The formation of a new phase of matter, a softer one than
nuclear matter,  may cause a rapidly rotating
pulsar to produce a prolonged signal
 that is dramatic,  easy to detect and easy to
understand \cite{glen97:a}.   The most plausible high
density phase  transition is deconfinement as predicted by
QCD \cite{asymptotic}.  The signal I will describe will
occur for  either a first or second order transition so long
as  it is accompanied by a sufficient softening of the
\eosp.  (Cf. Fig.\ \ref{eos}.)

Strictly speaking we do not even know that quarks can be
deconfined under  extreme conditions or otherwise. It is an
`expectation' based on the QCD property of asymptotic 
freedom \cite{asymptotic}. We would like to prove that this phase is a
possible  phase of matter. If so, it would have pervaded the
very early  universe, but  quark confinement
in hadrons
occurred at an early  time   and the thermal
equilibrium that existed then leaves no signal today.

\begin{figure}[htb]
\vspace{-.25in}
\begin{minipage}[t]{80mm}
\makebox[79mm]
{\psfig{figure=ps.qm1,width=3in,height=3.6in} }
\caption {The \eos (labeled `Hybrid') of
neutron star matter with a first order deconfinement phase transition.
The normal phase contains nucleons, hyperons and leptons in equilibrium.
The mixed phase contains as well, the three light flavor quarks.
Comparison is made with a case in which deconfinement is not taken account
of (labeled `n+p+H').
(Nuclear properties include the observed binding, saturation density,
symmetry energy and  $K=300$ MeV, $m^\star_{{\rm sat}}/m=0.7$)  
\label{eos_k300_y_h}
\label{eos}
}
\end{minipage}
\hspace{\fill}
\begin{minipage}[t]{75mm}
\makebox[74mm]
{\psfig{figure=ps.qm2,width=3in,height=3.6in}}
\caption { Density profiles
of two stars of the same mass $M=1.42 \msun$ but differing composition;
(1) Hyperon star (neutron-proton-hyperon-lepton), (2) Hybrid (a  
pure quark matter
core surrounded by mixed phase and outer pure hadronic
confined phase).  \Eoss as in Fig.\ \protect\ref{eos_k300_y_h}.
Interior differences are dramatic but not directly measureable.
(For a description of neutron star matter and
relativistic stars see Ref.\ \protect\cite{book}.
\label{prof_k240}
}
\end{minipage}
\end{figure}

From the balance of gravitational and centrifugal forces on
a particle at the surface of rapidly rotating stars such as
the millisecond pulsars, we know that the central density is
a few times nuclear, the same range of energy densities as
are expected to be produced in relativistic nuclear
collisions. Let us assume that the critical deconfinement
density occurs in the density range spanned by spherical
stars and therefore in the population of slow pulsars to
which the Crab belongs and of which there are about 800
presently known. In this case  newly born neutron stars have
a quark core essentially from birth. But we have no way to
tell if this is indeed the case. Models of neutron stars
having different 
composition  generally differ in the range of masses
and radii permitted, and their density profiles
may be very different (cf. Fig.\ \ref{prof_k240}). However as far as 
 measureable properties are concerned, ordinary neutron stars and
hybrid stars 
  (neutron stars with a quark core)
are 
practically indistinguishable. Cooling rates are presumeably sensitive
to the internal composition, but theoretical estimates are very
uncertain.

\begin{figure}[htb]
\vspace{-.25in}
\begin{center}
\begin{minipage}[t]{130mm}\hspace{.8in}
\makebox[79mm]
{\psfig{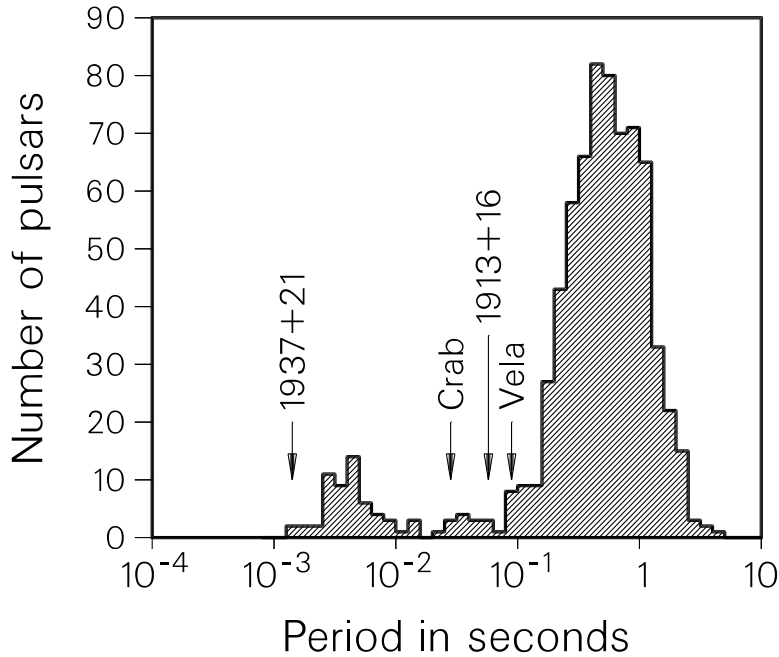} }
\caption { Distribution of pulsar periods. The lower group consists
of `recycled' millisecond pulsars, the higher to the canonical pulsars in the 
first stage of their evolution (see Fig.\ \protect\ref{pulsar}).
\label{period}
}
\end{minipage}
\end{center}
\vspace{-.35in}
\end{figure}

\begin{figure}[htb]
\vspace{-.25in}
\begin{center}
\begin{minipage}[t]{130mm}\hspace{.8in}
\makebox[79mm]
{\psfig{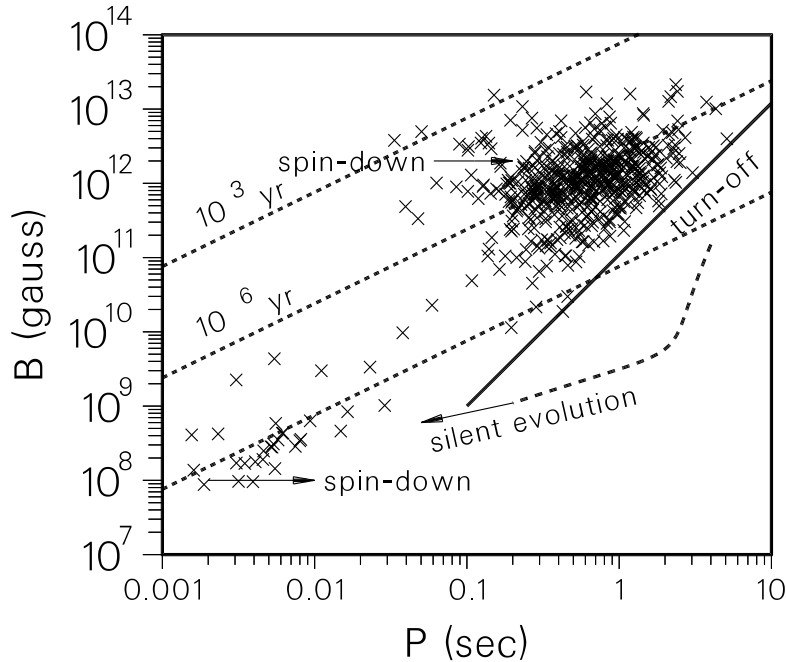} }
\vspace{-.35in}
\caption { Pulsars are born with field $B\sim 10^{12}
{\rm~to~} 10^{13}$ gauss and evolve
 toward the right. Periods change with time as $\sim 1/P$ and so 
pulsars accumulate
at large $P$. They become radio silent in 
about $10^7$ years and remain stagnant until they capture a companion star,
or unless they had one all along. Accretion from the less dense companion
spins them up along a line like `silent evolution'.  A combination of 
the now weaker field but higher frequency turns them on again as 
`recycled' millisecond
pulsars. They then again evolve toward shorter period, but now on a
very long time-scale because of the weaker field.
\label{pulsar}
}
\end{minipage}
\end{center}
\vspace{-.35in}
\end{figure}

I will sketch the generally accepted evolutionary  life
of pulsars
\cite{heuvel91:a}. There are two distinct populations of
pulsars (see Figs.\ \ref{period} and \ref{pulsar}),
the  canonical pulsars (about 800 of them now
known) with periods between  1/50 sec and 8 sec, and the
millisecond pulsars (about 50), which are believed  to be
more evolved than the former. Stars are born into the first
of these populations, and may, on a very long time-scale,
evolve into the second. (See Fig.\ \ref{pulsar}.)

As the stellar core of a luminous star collapses to form a 
neutron star, it is spun
up by conservation of angular momentum and acquires an
enormous magnetic field of $10^{12}{\rm~to~}10^{13}$ gauss
because of flux conservation. The
star is born as a rotating magnetic dipole. It has a
tremendous store of rotational energy that will keep it
spinning for 10 million years. The electromagnetic radiation
beamed along the spinning dipole is what we see as pulsed
radio emission once each rotation, if as observers, we lie on
the cone swept out by the beam. 

In a plot of magnetic field $B$ vs period $P$
(Fig.\ \ref{pulsar}), stars move
from top left to right because of loss of angular momentum to radiation. They disappear as active pulsars when
a combination of angular velocity  and field strength is
insufficient to produce radiation. It takes about $10^7$
years to complete this first phase. However, either from
birth, or afterward, the star may have  or capture  a lower
density companion as is often evidenced by the presence
of an orbiting white
dwarf. During an accretion era, the compact star is spun up by infalling
matter that it tears off from its less dense
companion. It
looses some magnetic field during accretion, perhaps by
ohmic resistance during the long radio silent  era. 

In this `silent' era the the star moves diagonally from top
right to bottom left in the $B-P$ diagram. The neutron star
becomes centrifugally flattened as it approaches millisecond
periods. The central density falls. The core of quark matter
shrinks as quarks recombine to form baryons. The quark core
may disappear altogether.

The silent neutron star, having completed a part of its life
cycle, turns on again as a millisecond pulsar of low
magnetic field when the lower field but higher angular
velocity can once again produce radiation. Presently 50
such pulsars have been discovered, half of which still have a
binary companion. The number of millisecond pulsars presently
known is believed to be a fraction
of the total population  because of
search selection effects. 

Millisecond  pulsars, which have    weaker fields ($10^8{\rm~to~}10^9$ gauss),
spin down very slowly since the deceleration is
proportional to $B^2$. Their characteristic age is $P/2\dot{P}\sim 10^9$ years.
The central density  is initially centrifugally diluted
but as it spins down, the central density will
rise again and the critical density will be reached, first
at the center, and then in an expanding region. The growth
of the central region of deconfined matter is paced by the
slow spin-down, slow because of the coupling of rotation of the  stellar 
magnetic dipole to electromagnetic processes.

Stiff  nuclear matter is being replaced in the core  by
highly compressible quark matter. The weight of the
overlaying layers of nuclear matter weigh down on the core
and compress it. Its density rises. The star shrinks---mass
is redistributed with growing concentration at the center.
The by-now more massive central region gravitationally
compresses the outer nuclear matter even further, amplifying
the effect. The density profile for a star at three angular
velocities,  (1) the limiting Kepler velocity which is stretched
in the equatorial plane and centrally diluted,  (2) an
intermediate angular velocity, and  (3) a non-rotating star,  are
shown in Fig.\  \ref{prof_k300b180}.
We see that the central density rises with
decreasing angular velocity by a factor of three and the
equatorial radius decreases by 30 percent.
In contrast, for a model for
which the phase transition did not take place, the central
density would change by only a few percent
\cite{weber90:d}.
The phase boundaries
are shown in Fig.\ \ref{omega_r_k300B180} from the highest
rotational frequency to zero rotation.
\begin{figure}[htb]
\vspace{-.5in}
\begin{minipage}[t]{80mm}
\makebox[79mm]
{\psfig{figure=ps.qm5,width=3in,height=3.6in} }
\caption { Mass profiles as a function of equatorial radius
of a star rotating at three different frequencies,
as marked. At low frequency the star is very dense in its core, having
a 4 km central region of highly compressible pure quark matter.
At intermediate frequency, the pure quark matter phase is absent and
the central 8 km is occupied by the mixed phase. At higher frequency (nearer $\Omega_K$) the star is relatively
dilute in the center and centrifugally stretched.  Inflections at
$\epsilon=220$ and $950$ are the boundaries of the mixed phase.
\label{prof_k300b180}
}
\end{minipage}
\hspace{\fill}
\begin{minipage}[t]{75mm}
\makebox[74mm]
{\psfig{figure=ps.qm6,width=3in,height=3.6in} }
\caption {  Radial boundaries at various rotational frequencies
separating  (1) pure quark matter,  (2)
mixed phase, (3) pure hadronic phase, (4) ionic crust of neutron 
rich nuclei and surface of star. The pure
quark phase appears only when the  frequency is below $\Omega
\sim 1370$ rad/s. Note the decreasing radius as the frequency falls.
The frequencies of two pulsars, the Crab and
PSR 1937+21 are marked for reference.
\label{omega_r_k300B180}
}
\end{minipage}
\end{figure}
The redistribution of mass and shrinkage of the star change
its moment of inertia and hence the characteristics of its
spin behavior. The star must spin up to conserve angular momentum
which is being carried off only  slowly by the weak electromagnetic
dipole radiation. The star  behaves like an ice skater
who goes into a spin with arms outstretched, is slowly spun
down by friction, temporarily spins up by pulling the arms
inward, after which friction takes over again.

It is that simple to describe, and that is the blatant signal I 
mentioned---the spontaneous spin-up of an isolated millisecond pulsar 
that is 
radiating angular momentum and ought otherwise to be slowing down.

\section{BACKBENDING IN PULSARS: Signal of Phase Transition}

The mathematical description of the intuitive process
described above is  difficult but well
defined. We must compute in \GR a sequence of rotating
stellar objects of the same baryon number, but varying in
angular velocity based on an \eos
that describes a phase transition as in
Fig.\ \ref{eos_k300_y_h}. The expression for the  moment of inertia
of a non-rotating star
was derived years ago by Hartle and Thorne but it is inadequate
\cite{hartle67:a,hartle68:a}. It
ignores the centrifugal stretching of the star, changes in
its composition that result, changes in the metric of space
time as a result of rotation,  and it ignores the dragging
of local inertial frames by the rotating star. 
Fortunately, in a different connection, we derived the General
Relativistic
expression for a rotating star that incorporates all of the above
effects \cite{glen92:b,glen93:a}.
\begin{figure}[htb] 
\vspace{-1.75in} \begin{center}
\begin{minipage}[t]{130mm}   
{\psfig{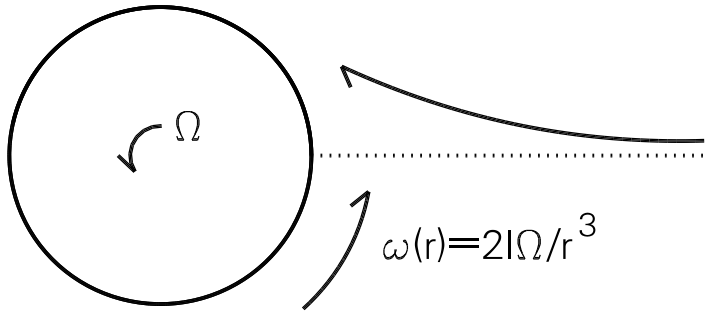} }
\vspace{-.75in} \caption { A star rotating with angular
velocity $\Omega$ as measured by a distant observer, sets
the local inertial frames into rotation by a position
dependent angular velocity $\omega(r)$. A particle  dropped
on the star in the equatorial plane falls not toward the
center, but is swept to the side as illustrated.  Outside the star
$\omega$ has a simple form.
\label{circle} }
\end{minipage} 
\end{center}
\vspace{-.25in} 
\end{figure}

\begin{figure}[htb]
\vspace{-1in}
\begin{center}
\begin{minipage}[t]{130mm} \hspace{.5in}
\makebox[79mm]
{\psfig{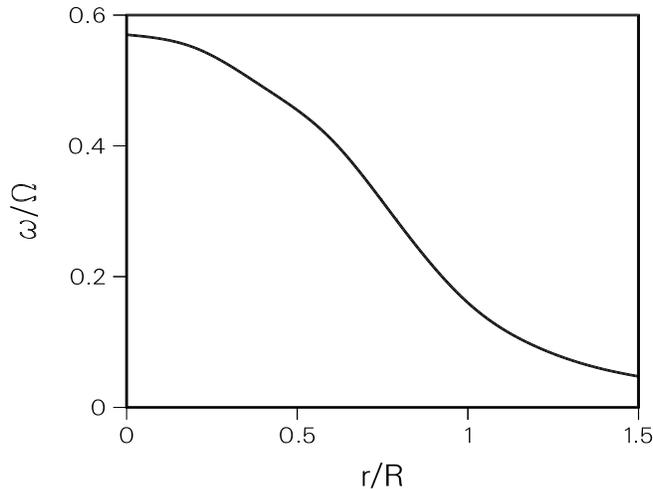} }
\caption { Angular velocity $\omega(r,\theta)$
 of local inertial frames as a fraction of the
star's angular velocity $\Omega$ is shown as a function of radial coordinate
in the equatorial plane of rotation.
The centrifugal forces are determined not by $\Omega$ but by 
$\Omega-\omega(r,\theta)$. Inside the star, $\omega(r,\theta)$ influences 
and in turn is influenced by the distribution of matter.
\label{drag}
}
\end{minipage}
\end{center}
\vspace{-.25in}
\end{figure}

Frame dragging by a rotating star is as inseparable from a
description of space-time as mass is \cite{book}. A particle
dropped  in the equatorial plane
onto a rotating star falls not toward the star's
center, but is swept ever more in the sense of the stars
rotation, as illustrated in Fig.\ \ref{circle}. The
centrifugal force acting on a fluid element of the star is
governed, not by its angular velocity $\Omega$, but by the
angular velocity relative to that of the local inertial
frames $\omega(r)$. The latter is a function of position as
shown in Fig.\ \ref{drag}.

For a sequence of stars near the mass limit, we show in
Fig.\ \ref{oi} the moment of inertia in a small band of
angular velocity  in the vicinity where the quark matter core
radius grows rapidly  with respect to changing $\Omega$ (cf.
Fig.\ \ref{omega_r_k300B180}). Normally, the moment of inertia
would decrease as a smooth function of  $\Omega$ as the
centrifugal force allows the star to relax to a  more nearly
spherical shape. Instead, we see the behavior just described
for an ice skater. The neutron star actually spins up during
the growth of the quark core. The reason, as explained, is
that as stiff nuclear matter is replaced by compressible
quark matter, the  moment of inertia decreases,
not solely because of the diminishing centrifugal forces, but
because of the change of state; the star
must spin up to conserve angular momentum. A very similar
phenomenon was predicted for rotating nuclei in 1960 by Mottelson and Valatin
and observed in the 1970's
\cite{mottelson60:a,johnson72:a,stephens72:a}. A  nucleus 
undergoes a phase change from a phase at high angular
velocity in which the nucleus carries its angular momentum
through aligned  nucleon spins, to a phase at  low angular
momentum which is superfluid. The change takes place over a
few units of angular velocity and the ``backbending'' of the
moment of inertia is shown in Fig. \ref{nucleus}.
\begin{figure}[htb] 
\vspace{-.35in}
\begin{minipage}[t]{80mm} 
\makebox[79mm]
{\psfig{figure=ps.qm9,width=3in,height=3.6in} }
\caption { Moment of inertia of a neutron star at angular
velocities for which the central density rises from below to
above critical density for the pure quark matter phase as
the centrifugal force decreases. Time flows from large to small
$I$. The most arresting signal
of the phase change is the spontaneous spin-up that  an
isolated pulsar would undergo during the growth in the
region of pure quark matter.   \label{oi} } 
\end{minipage} %
\hspace{\fill} %
\begin{minipage}[t]{75mm} 
\makebox[74mm]
{\psfig{figure=ps.qm10,width=3in,height=3.6in} }
\caption {  Nuclear moment of inertia as a function of squared 
frequency for $^{158}$Er, showing backbending in the nuclear case.
Quantization of spin yields the unsmooth curve compared to 
the one in Fig.\
\protect\ref{oi}. There the spin at the center of the spin-up is
$J\sim 10^{41}$.
\label{nucleus}
}
\end{minipage}
\end{figure}

It is trivial to measure the rate at which a pulsar's period changes.
It is especially trivial for millisecond pulsars
where the period is sometimes known to 14 significant
figures. For example, PSR1937+21 has a period (measured on 29 November 1982
at 1903 UT)
$$P= 1.5578064487275(3) {\rm~ms}\,.$$
Its rate of change of period is a mere $\dot{P}\sim 10^{-19}$.
But because of the high accuracy of the period measurement, it takes only
two measurements spaced 
$0.3$  hours apart to detect a unit change in the last significant
figure, and hence to detect in which direction the period is changing. 
So in fact we have uncovered a signal of a phase
transition in pulsars that is trivial to observe if it
occurs during the time of observation. Let us now compute
the length of the epoch over which the pulsar will be
spinning up because of a change of phase and the  slow 
envelopment of the central region by the new phase.

\section{DURATION OF THE SPIN-UP ERA}
Having the moment of inertia as a function of 
angular velocity  for a model star, as in
Fig.\ \ref{oi}, that is to say, for a model of the 
 \eos which describes a phase transition, we are in a position to compute 
the time evolution of the angular velocity. The rate at which energy is 
radiated by a rotating magnetic dipole is give by 
 \begin{eqnarray}
 \frac{dE}{dt} =
  \frac{d}{dt}\Bigl(\frac{1}{2} I \Omega^2 \Bigr) =
   - C \Omega^{4}\,.
    \label{energyloss}
      \end{eqnarray}
Here, $I$ is the moment of inertia, $\Omega$ is the angular
velocity of the star and $C$ is a constant that depends on
the square of the  magnetic field strength. Therefore, the rate of change
of frequency is governed by \begin{eqnarray} \dot{\Omega}=
-\frac{C}{I(\Omega)}
  \biggl[1  + \frac{I^{\prime}(\Omega) \,
\Omega}{2I(\Omega)}
 \biggr]^{-1} \Omega^3
   \label{braking2}
\end{eqnarray}
where $I^\prime =dI/d\Omega$.
From the behavior of the moment of inertia, 
we see that its derivative is infinite at the two frequencies
that mark  the beginning and end of the 
spin-up era; consequently $\dot{\Omega}$ vanishes at the boundaries
of this era.

By integrating the equation we can find the length of the
epoch of spin-up. It is $2\times 10^7$ years. The
characteristic time for spin-down of millisecond pulsars is
$10^9$ years. So if the conditions for spin-up are
fulfilled, namely (i) the equation of state describes a
phase transition involving a substantial softening of the
\eosp, and (ii) the critical density is attained in stars
very near the mass limit,
we can expect a spontaneous spin-up of millisecond
pulsars
to occur  in the above ratio, namely 1/50. This is a very
attractive ``event rate'' given that 25 of the presently
known millisecond pulsars are isolated.

There is another observable which endures for an even longer
time. It is the so-called  dimensionless braking index
 ${\Omega \ddot{\Omega} }/{\dot{\Omega}^2}$.
It would be  equal to the magnetic dipole value $n=3$
 of the energy-loss
 mechanism (\ref{energyloss})
 if the frequency were small or
if the moment of inertia were a constant.
However these conditions are not usually fulfilled and the measurable
quantity is not  constant. Rather it has the value
  \begin{eqnarray}
  n(\Omega)\equiv\frac{\Omega \ddot{\Omega} }{\dot{\Omega}^2}
  = 3   - \frac{ 3  I^\prime \Omega +I^{\prime \prime} \Omega^2 }
    {2I + I^\prime \Omega}
     \label{index}
      \end{eqnarray}
      where  $I^{\prime\prime}
      = dI^2/d\Omega^2$.
\begin{figure}[tbh]
\vspace{-.5in}
\begin{center}
\begin{minipage}[t]{130mm} \hspace{.75in}
\makebox[79mm]
{\psfig{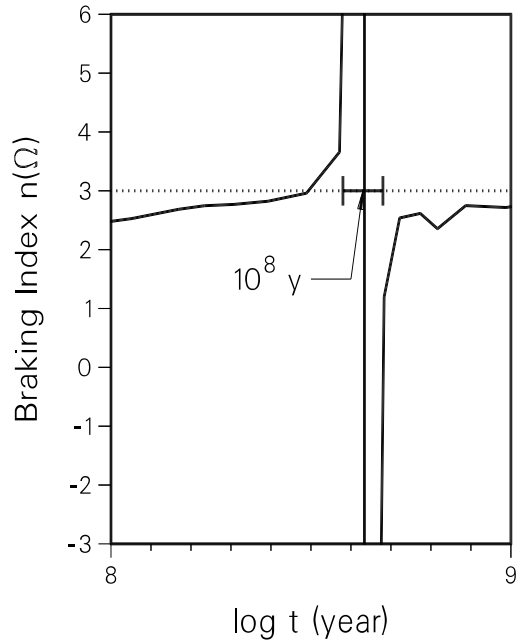} }
\caption { The time evolution of
the braking index plotted over one
decade that includes the epoch of the phase transition. The singularities
correspond to the infinite derivatives of $dI/d\Omega$ that bound the
spin-up era. (See Fig.\ \protect\ref{oi} and the denominator of Eq.
(\protect\ref{braking2}).) The line connecting
the singularities corresponds to
the spin-up era.
\label{nt}
}
\end{minipage}
\end{center}
\vspace{-.35in}
\end{figure}
The progression of the new phase through the central region
of the star will be signaled by an anomalous value of the
braking index, far removed from the canonical value of $3$.
Indeed, at the two turning points of $I$ that mark the boundaries of the
spin-up era, $\dot{\Omega}$ vanishes so that $n(\Omega)$ is infinite and of
opposite sign. 
We show the behavior of $n(\Omega)$ in Fig. \ref{nt}.
The epoch over
which it is anomalous is $10^8$ years.    Note that one
cannot, and does not need to measure the shape of the curve.
A single anomalous value that differed significantly from
the dipole value of 3 would suffice. How could one reconcile a value
of the multipolarity of say 10, or $-5$ with observed electromagnetic
processes?  However it is hard to
measure the second time derivative and therefore the braking
index, especially for
millisecond pulsars. So the practical signal is spontaneous
spin-up of an isolated pulsar. We specify `isolated' to ensure that spin-up
is not attributable to accretion from a companion.

\section{SUMMARY}
An isolated millisecond pulsar
will spin up
over an epoch of $2 \times10^7$  years out of a  spin-down life of
$10^9$  years if it undergoes a phase
transition obeying the two conditions 
(i) the  transition causes a substantial
softening of the
\eosp, and (ii) the critical density is attained in stars
very near the mass limit.
The spin-up epoch, compared to the spin-down life of the pulsar,
 corresponds to an `event rate' of 1/50.
The determination of whether a pulsar is spinning up or down
is trivial. Of the presently known millisecond pulsars,
about 25 are isolated. We are approaching the moment of
truth for this observable signal of a phase transition.

We have emphasized that the transition need not be of first
order so long as it is accompanied by a sufficient softening
of the \eosp. We do not have a measure of what we mean by
this. Of course our model does possess the requisite
softening, or our results would not have exhibited
backbending. 

If no pulsar is observed to produce the signal, little is
learned. Just as in the search for deconfinement
in high energy  nuclear collisions, failure to observe a
signal does not inform us that the deconfined phase does not
exist.


\end{document}